\documentclass[12pt]{article}
\usepackage{graphicx}
\def \beq{\begin{equation}}
\def \eeq{\end{equation}}
\def \ite{{\it et al.}}

\begin{document}
\rightline{EFI 06-23}
\rightline{hep-ph/0611207}
\bigskip

\centerline{\bf MASS SPLITTINGS IN $\Sigma_b$ and $\Sigma^*_b$
\footnote{To be published in Phys.~Rev.~D, Brief Reports.}}
\bigskip
\centerline{\it Jonathan L. Rosner}
\centerline{\it Enrico Fermi Institute and Department of Physics,
University of Chicago}
\centerline{\it 5640 S. Ellis Avenue, Chicago IL 60615}
\bigskip

\centerline{\bf ABSTRACT}
\bigskip

\begin{quote}
The charged $\Sigma_b$ and $\Sigma^*_b$ states have recently been reported by
the CDF Collaboration.  The relation of their reported charge-averaged masses
to expectations based on the quark model is reviewed briefly.
A relation is proved among the $\Delta I = 1$ electromagnetic mass differences
$\Sigma_1 \equiv M(\Sigma^+) - M(\Sigma^-)$, $\Sigma^*_1 \equiv M(\Sigma^{*+})-
M(\Sigma^{*-})$, $\Sigma_{b1} \equiv M(\Sigma_b^+) - M(\Sigma_b^-)$, and
$\Sigma^*_{b1} \equiv M(\Sigma_b^{*+}) - M(\Sigma_b^{*-})$.  The relation
is $\Sigma^*_{b1} - \Sigma_{b1} = (m_s/m_b)(\Sigma^*_1 - \Sigma_1),$ leading
to the expectation $\Sigma^*_{b1} - \Sigma_{b1} = 0.40 \pm 0.07$ MeV.
\end{quote}

The Collider Detector Facility (CDF) Collaboration at Fermilab has recently
announced the observation of four new candidates for $\Sigma_b^\pm$ and
$\Sigma_b^{*\pm}$ \cite{CDFsigb}, with masses very close to those expected in
theory.  Ref.\ \cite{Jenkins:1996de} uses a double expansion in $1/N_c$ and
$1/m_Q$, where $N_c$ is the number of quark colors and $m_Q$ is the heavy quark
mass, while Ref.\ \cite{Karliner:2003sy} uses the quark model.  A recent
relativistic calculation and comparison with some earlier predictions
\cite{earlier} may be found in Ref.\ \cite{Ebert:2005xj}.

\begin{figure}
\begin{center}
\includegraphics[height=4.5in]{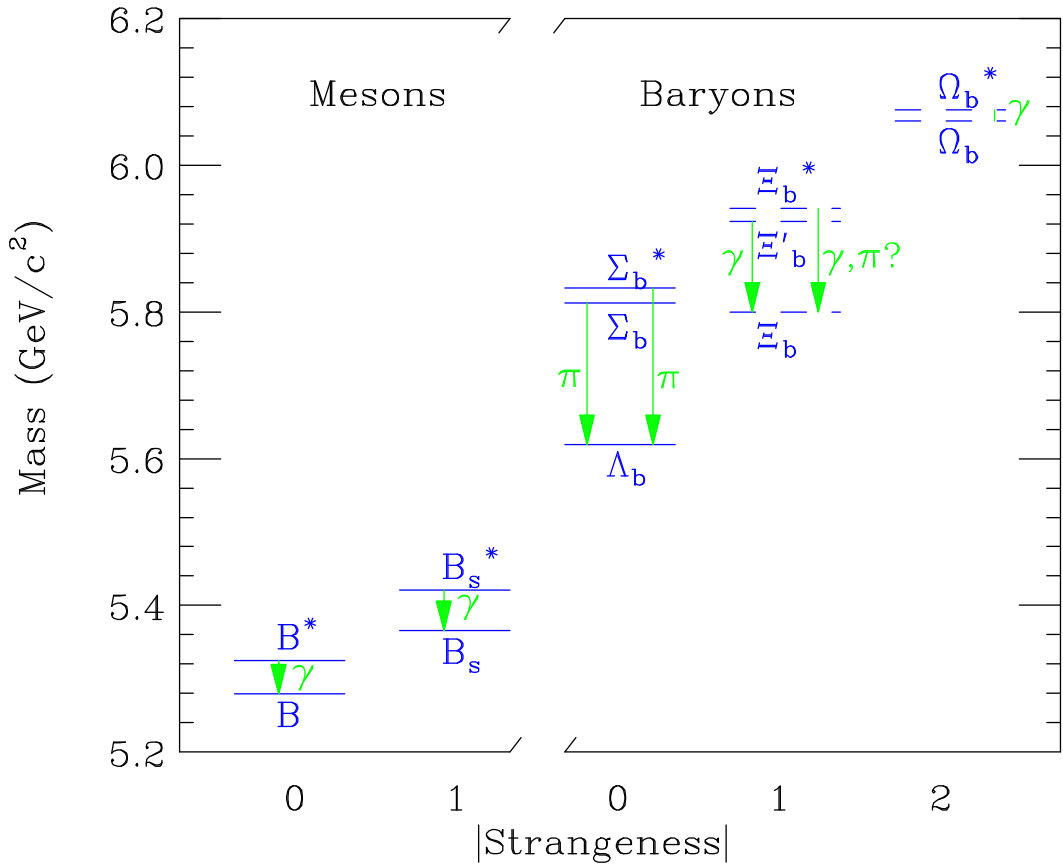}
\end{center}
\caption{Hadrons containing a single beauty quark.  The $\Sigma_b$ and
$\Sigma^*_b$ have recently been reported by CDF \cite{CDFsigb}.  Dashed lines
denote states not yet observed.
\label{fig:beauty}}
\end{figure}

The $\Sigma_b$ and $\Sigma^*_b$ states are illustrated in Fig.\
\ref{fig:beauty}.  They would have quark content $buu,~bdd$ with total spins
$J(\Sigma_b^\pm) = 1/2$ and $J(\Sigma_b^{*\pm}) = 3/2$.

The analysis of Ref.\ \cite{CDFsigb} studies the spectra of $\Lambda_b \pi^\pm$
states, finding peaks at the values of $Q^{(*)\pm} \equiv M(\Sigma^{(*)\pm})
- M(\pi^\pm) - M(\Lambda_b)$ shown in Table \ref{tab:sigb}.  These may be
combined with the newly reported CDF value $M(\Lambda_b) = 5619.7 \pm 1.7 \pm
1.7$ MeV \cite{Acosta:2005mq} to obtain masses of the $\Sigma_b^{(*)\pm}$
states.  Here $Q$ and $Q^*$ denote the averages of $Q^\pm$ and $Q^{*\pm}$,
respectively.  In this analysis it was assumed that $Q^{*+} - Q^{*-} =
Q^+ - Q^-$.  The main point of the present paper is to examine the validity
of this assumption.

\begin{table}[h]
\caption{Values of $Q^{(*)\pm} \equiv M(\Sigma_b^{(*)\pm}) - M(\pi^\pm) -
M(\Lambda_b)$ and $M(\Sigma^{(*)\pm})$ reported by the CDF Collaboration
\cite{CDFsigb}.
\label{tab:sigb}}
\begin{center}
\begin{tabular}{c c} \hline \hline
Quantity & Value (MeV) \\ \hline
$Q^+$ & $48.4^{+2.0}_{-2.3} \pm 0.1$ \\
$Q^-$ & $55.9 \pm 1.0 \pm 0.1$ \\
$Q^* - Q$ & $21.3^{+2.0+0.4}_{-1.9-0.2}$ \\ \hline \hline
\end{tabular}
\end{center}
\end{table}

We begin by discussing the non-electromagnetic mass splittings briefly.  Here
the basic physics is the same as that in \cite{Gasiorowicz:1981jz}, which may
be consulted for earlier references.  The charge-averaged hyperfine splitting
between the $J=1/2$ and $J=3/2$ states may be predicted from that for charmed
particles:
\beq
\frac{M(\Sigma_b^*) - M(\Sigma_b)}{M(\Sigma_c^*) - M(\Sigma_c)} =
\frac{m_c}{m_b} = \frac{1.5~{\rm GeV}}{4.9~{\rm GeV}} = 0.31~~,
\eeq
where we use ``constituent'' quark masses from Ref.\ \cite{Kwong:1987ak}.
Using isospin-averaged differences $M(\Sigma_c) - M(\Lambda_c) = (167.09
\pm 0.13)$ MeV and $M(\Sigma^*_c) - M(\Lambda_c) = (231.5 \pm 0.8)$ MeV
based on Ref.\ \cite{Yao:2006px}, we find this ratio to be $0.33 \pm 0.03$.
The first of Refs.\ \cite{Jenkins:1996de} finds $M(\Sigma_b^*) - M(\Sigma_b) =
23.8$ MeV, the second finds 15.8 MeV, and Ref.\ \cite{Ebert:2005xj} finds 29
MeV.

As for the splitting of the spin-weighted average $[2 M(\Sigma_b^*) +
M(\Sigma_b)]/3$ from the $\Lambda_b$, it is expected to be the same as
the corresponding values for hyperons containing strange or charmed quarks.
The experimental values are \cite{CDFsigb}
\beq
\frac{M(\Sigma_b) + 2 M(\Sigma_b^*)}{3} - M(\Lambda_b) = (205.9 \pm 1.8)~{\rm
MeV},~
\eeq
where we have used the averages of the differences for $\Sigma_b^{(*)\pm}$
as no information is available on $M(\Sigma_b^{(*)0})$.
This is to be compared with
\beq
\frac{M(\Sigma_c) + 2 M(\Sigma_c^*)}{3} - M(\Lambda_c) = (210.0 \pm 0.5)~{\rm
 MeV},
\eeq
where we have used the differences with respect to $M(\Lambda_c)$ mentioned
above, and
\beq
\frac{M(\Sigma) + 2 M(\Sigma^*)}{3} - M(\Lambda) = (205.1 \pm 0.3)~{\rm MeV}~,
\eeq
where the masses are taken directly from Ref.\ \cite{Yao:2006px}, and an
average over the $\Sigma$ isospin multiplet is taken.  In each case the
dominant source of error is the mass of the $I_3 = 0$, $J=3/2$ state,
$\Sigma_c^{*+}$ or $\Sigma^{*0}$.

Ref.\ \cite{Jenkins:1996de} also predicts the equality of these mass
splittings, and estimates that the $b$ and $c$ quantities should be equal to
about $\pm 5$ MeV.  Ref.\ \cite{Karliner:2003sy} uses quark-model arguments
to estimate the $\Sigma_b$ mass but eliminates reference to actual quark masses
by using other hadron mass splittings.
\medskip

We now turn to electromagnetic mass splittings.
The discussion will be conducted in a quark model in which there are several
sources of baryon electromagnetic mass differences \cite{Rosner:1998zc}.  Most
of these cancel out when one takes the $\Delta I = 1$ mass differences
\beq
\Sigma_1 \equiv M(\Sigma^+) - M(\Sigma^-)~,~~
\Sigma^*_1 \equiv M(\Sigma^{*+})- M(\Sigma^{*-})~,~~
\eeq
\beq
\Sigma_{b1} \equiv M(\Sigma_b^+) - M(\Sigma_b^-)~,~~
\Sigma^*_{b1} \equiv M(\Sigma_b^{*+}) - M(\Sigma_b^{*-})~.
\eeq
However, we review briefly all sources of isospin violation in baryon masses.
\medskip

\noindent
{\it 1.  Intrinsic quark masses.}

The $u$ and $d$ quarks have intrinsic masses which differ by a couple of MeV
\cite{Yao:2006px}.  Corresponding estimates for the strange quark mass are in
the vicinity of 100 MeV.  However, quarks in hadrons are more suitably
described by the ``constituent'' values (see, e.g., Refs.\
\cite{Gasiorowicz:1981jz} and \cite{DeRujula:1975ge}) $m_u,m_d = {\cal O}(350)$
MeV, $m_s = {\cal O}(500)$ MeV, with $m_d - m_u$ of order a few MeV but quite
uncertain.  The quarks' kinetic energies may also depend on their masses.
Without detailed knowledge of dynamics, it is difficult to anticipate this
dependence.  One may simply parametrize kinetic energies with labels $K_q$ for
those contributions which act as one-body operators and $K_{q_i q_j}$ for those
contributions which depend on interactions with each individual other quark.
\medskip

\noindent
{\it 2.  Coulomb interactions between quarks.}

Each quark pair in a hadron has a Coulomb interaction energy
\beq \label{eqn:coul}
\Delta E_{ij~\rm em} = \alpha Q_i Q_j \langle \frac{1}{r_{ij}}
\rangle~~~,
\eeq
where $\alpha$ is the electromagnetic fine structure constant, $Q_i$ is the
charge of quark $i$ in units of the proton charge, and $\langle 1/r_{ij}
\rangle$ is the expectation value of the inverse distance between the members
of the pair.  In the flavor-SU(3) limit $\langle 1/r_{ij} \rangle$ will be
universal throughout a multiplet.  In this limit, we parametrize the
interaction energy $\Delta E_{ij~\rm em} = a Q_i Q_j$, where
$a$ is some universal constant.
\medskip

\noindent
{\it 3.  Strong hyperfine interactions.}

Quarks in hadrons experience a spin-dependent force due to gluon exchange which
acts dominantly on pairs in an S-wave state.  For quark pairs in a baryon, one
has a strong hyperfine interaction energy 
\beq \label{eqn:HFs}
\Delta E_{ij~\rm HFs} = {\rm const.} \frac{|\Psi_{ij}(0)|^2
\langle \sigma_i \cdot \sigma_j \rangle}{m_i m_j}~~~,
\eeq
where $|\Psi_{ij}(0)|^2$ is the square of the S-wave wave function of two
quarks at zero relative separation, and the constant is universal for all
quark pairs in a baryon.  We shall assume that $|\Psi_{ij}(0)|^2$ is universal
for all quark pairs in S-wave baryons.  We then find a contribution to the
hyperfine energy $\Delta E_{ij~\rm HFs} = \beta \langle \sigma_i \cdot \sigma_j
\rangle/ (m_i m_j)$.  The calculation of strong hyperfine splittings in
baryons, requiring evaluation of $\langle \sigma_i \cdot \sigma_j \rangle$ for
each quark pair, is described in more detail in Ref.\ \cite{Rosner:1998zc}.
\medskip

\noindent
{\it 4.  Electromagnetic hyperfine interactions.}

The electromagnetic interaction between quarks in a baryon has a
hyperfine contribution
\beq \label{eqn:HFe}
\Delta E_{ij~\rm HFe} = - \frac{2 \pi \alpha Q_i Q_j |\Psi(0)_{ij}|^2
\langle \sigma_i \cdot \sigma_j \rangle}{3 m_i m_j}~~~.
\eeq
Assuming universality of the wave functions, we parametrize this effect as
$\Delta E_{ij~\rm HFe} = \gamma Q_i Q_j \langle \sigma_i \cdot \sigma_j
\rangle/(m_i m_j)$.
\medskip

We now form the {\it differences} of $\Delta I = 1$ mass differences
for $\Sigma^\pm$ and $\Sigma^{*\pm}$ states.  We find
\beq \label{eqn:sigdif}
\Sigma^*_1 - \Sigma_1 = \beta \left(\frac{6}{m_u m_s} - \frac{6} {m_d m_s}
\right) - \frac{\gamma}{9} \left(\frac{6}{m_d m_s} + \frac{12}{m_u m_s} \right)
= - 2 \sqrt{3} M_{\Lambda \Sigma^0}~~~,
\eeq
where $M_{\Lambda \Sigma^0}$ is an isospin-violating term mixing the $\Lambda$
and $\Sigma^0$.  Corresponding relations may be written for $\Sigma_b$ and
$\Sigma^*_b$ with the substitution $s \to b$:
\beq \label{eqn:sigbdif}
\Sigma^*_{b1} - \Sigma_{b1} = \beta \left(\frac{6}{m_u m_b} - \frac{6}{m_d m_b}
\right) - \frac{\gamma}{9} \left(\frac{6}{m_d m_b} + \frac{12}{m_u m_b} \right)
= - 2 \sqrt{3} M_{\Lambda_b \Sigma_b^0}~~~.
\eeq
The crucial point is that these differences are of order $1/m_s$ and $1/m_b$,
respectively.  They are thus related by
\beq \label{eqn:reln}
\Sigma^*_{b1} - \Sigma_{b1} = (m_s/m_b)(\Sigma^*_1 - \Sigma_1)~~~.
\eeq
This relation is implicit in many previous treatments (see, e.g., Table I
in the first of Refs.\ \cite{Jenkins:1996de}), as any such hyperfine
differences are expected to scale as the inverse of the heavy quark mass and
should be the same for quarks of the same charge ($s$ and $b$ in the present
case).

We now use the experimental averages \cite{Yao:2006px}
\beq
\Sigma_1 = - 8.08 \pm 0.08 ~{\rm MeV}~,~~
\Sigma_1^* = -4.4 \pm 0.64  ~{\rm MeV}~,~~
\eeq
and the constituent-quark masses $m_s = 538$ MeV \cite{Gasiorowicz:1981jz}
and $m_b = 4.9$ GeV \cite{Kwong:1987ak} to predict
\beq \label{eqn:sigbdifval}
\Sigma^*_{b1} - \Sigma_{b1} = 0.40 \pm 0.07~{\rm MeV}
\eeq
While the analysis of Ref.\ \cite{CDFsigb} assumed $\Sigma^*_{b1} - \Sigma_{b1}
= 0$ (which would be accurate in the limit of $m_b \to \infty$), the relatively
small value in Eq.\ (\ref{eqn:sigbdifval}) is not likely to lead to a
substantial change in masses obtained from experiment.
Earlier discussions of electromagnetic mass splittings in heavy baryons also
find small values of $\Sigma^*_{b1} - \Sigma_{b1}$, without explicitly noting
the relation (\ref{eqn:reln}): 0.6 MeV in \cite{Hwang:1986ee} and 0.2 MeV in
\cite{Capstick:1987cw}. 

A corresponding relation cannot be obtained for the charmed baryon mass
differences $\Sigma_{c1} \equiv M(\Sigma_c^{++}) - M(\Sigma_c^0)$ and
$\Sigma^*_{c1} \equiv M(\Sigma_c^{*++}) - M(\Sigma_c^{*0})$.  All that can be
said is that in the limit of $m_c \to \infty$, one would have $\Sigma^*_{c1} -
\Sigma_{c1} \to 0$.  Present data give $\Sigma_{c1} = (0.27 \pm 0.11)$ MeV
and $\Sigma^*_{c1} = (0.3 \pm 0.6)$ MeV \cite{Yao:2006px}.
\bigskip

I thank R. Faustov, B. Grinstein, I. Gorelov, M. Karliner, G. Paz, and
S. Narison for useful communications.  This work was supported in part by the
United States Department of Energy under Grant No. DE FG02 90ER40560. 
\newpage

\def \ajp#1#2#3{Am. J. Phys. {\bf#1}, #2 (#3)}
\def \apny#1#2#3{Ann. Phys. (N.Y.) {\bf#1}, #2 (#3)}
\def \app#1#2#3{Acta Phys. Polonica {\bf#1}, #2 (#3)}
\def \arnps#1#2#3{Ann. Rev. Nucl. Part. Sci. {\bf#1}, #2 (#3)}
\def \cmts#1#2#3{Comments on Nucl. Part. Phys. {\bf#1}, #2 (#3)}
\def \cn{Collaboration}
\def \cp89{{\it CP Violation,} edited by C. Jarlskog (World Scientific,
Singapore, 1989)}
\def \dpfa{{\it The Albuquerque Meeting: DPF 94} (Division of Particles and
Fields Meeting, American Physical Society, Albuquerque, NM, Aug.~2--6, 1994),
ed. by S. Seidel (World Scientific, River Edge, NJ, 1995)}
\def \dpff{{\it The Fermilab Meeting: DPF 92} (Division of Particles and Fields
Meeting, American Physical Society, Batavia, IL., Nov.~11--14, 1992), ed. by
C. H. Albright \ite~(World Scientific, Singapore, 1993)}
\def \efi{Enrico Fermi Institute Report No. EFI}
\def \epl#1#2#3{Europhys.~Lett.~{\bf #1}, #2 (#3)}
\def \f79{{\it Proceedings of the 1979 International Symposium on Lepton and
Photon Interactions at High Energies,} Fermilab, August 23-29, 1979, ed. by
T. B. W. Kirk and H. D. I. Abarbanel (Fermi National Accelerator Laboratory,
Batavia, IL, 1979}
\def \hb87{{\it Proceeding of the 1987 International Symposium on Lepton and
Photon Interactions at High Energies,} Hamburg, 1987, ed. by W. Bartel
and R. R\"uckl (Nucl. Phys. B, Proc. Suppl., vol. 3) (North-Holland,
Amsterdam, 1988)}
\def \ib{{\it ibid.}~}
\def \ibj#1#2#3{~{\bf#1}, #2 (#3)}
\def \ichep72{{\it Proceedings of the XVI International Conference on High
Energy Physics}, Chicago and Batavia, Illinois, Sept. 6 -- 13, 1972,
edited by J. D. Jackson, A. Roberts, and R. Donaldson (Fermilab, Batavia,
IL, 1972)}
\def \ijmpa#1#2#3{Int. J. Mod. Phys. A {\bf#1}, #2 (#3)}
\def \jpb#1#2#3{J.~Phys.~B~{\bf#1}, #2 (#3)}
\def \lkl87{{\it Selected Topics in Electroweak Interactions} (Proceedings of
the Second Lake Louise Institute on New Frontiers in Particle Physics, 15 --
21 February, 1987), edited by J. M. Cameron \ite~(World Scientific, Singapore,
1987)}
\def \ky85{{\it Proceedings of the International Symposium on Lepton and
Photon Interactions at High Energy,} Kyoto, Aug.~19-24, 1985, edited by M.
Konuma and K. Takahashi (Kyoto Univ., Kyoto, 1985)}
\def \mpla#1#2#3{Mod. Phys. Lett. A {\bf#1}, #2 (#3)}
\def \nc#1#2#3{Nuovo Cim. {\bf#1}, #2 (#3)}
\def \np#1#2#3{Nucl. Phys. {\bf#1}, #2 (#3)}
\def \pisma#1#2#3#4{Pis'ma Zh. Eksp. Teor. Fiz. {\bf#1}, #2 (#3) [JETP Lett.
{\bf#1}, #4 (#3)]}
\def \pl#1#2#3{Phys. Lett. {\bf#1}, #2 (#3)}
\def \pla#1#2#3{Phys. Lett. A {\bf#1}, #2 (#3)}
\def \plb#1#2#3{Phys. Lett. B {\bf#1}, #2 (#3)}
\def \pr#1#2#3{Phys. Rev. {\bf#1}, #2 (#3)}
\def \prc#1#2#3{Phys. Rev. C {\bf#1}, #2 (#3)}
\def \prd#1#2#3{Phys. Rev. D {\bf#1}, #2 (#3)}
\def \prl#1#2#3{Phys. Rev. Lett. {\bf#1}, #2 (#3)}
\def \prp#1#2#3{Phys. Rep. {\bf#1}, #2 (#3)}
\def \ptp#1#2#3{Prog. Theor. Phys. {\bf#1}, #2 (#3)}
\def \ptwaw{Plenary talk, XXVIII International Conference on High Energy
Physics, Warsaw, July 25--31, 1996}
\def \rmp#1#2#3{Rev. Mod. Phys. {\bf#1}, #2 (#3)}
\def \rp#1{~~~~~\ldots\ldots{\rm rp~}{#1}~~~~~}
\def \si90{25th International Conference on High Energy Physics, Singapore,
Aug. 2-8, 1990}
\def \slc87{{\it Proceedings of the Salt Lake City Meeting} (Division of
Particles and Fields, American Physical Society, Salt Lake City, Utah, 1987),
ed. by C. DeTar and J. S. Ball (World Scientific, Singapore, 1987)}
\def \slac89{{\it Proceedings of the XIVth International Symposium on
Lepton and Photon Interactions,} Stanford, California, 1989, edited by M.
Riordan (World Scientific, Singapore, 1990)}
\def \smass82{{\it Proceedings of the 1982 DPF Summer Study on Elementary
Particle Physics and Future Facilities}, Snowmass, Colorado, edited by R.
Donaldson, R. Gustafson, and F. Paige (World Scientific, Singapore, 1982)}
\def \smass90{{\it Research Directions for the Decade} (Proceedings of the
1990 Summer Study on High Energy Physics, June 25--July 13, Snowmass, Colorado),
edited by E. L. Berger (World Scientific, Singapore, 1992)}
\def \tasi90{{\it Testing the Standard Model} (Proceedings of the 1990
Theoretical Advanced Study Institute in Elementary Particle Physics, Boulder,
Colorado, 3--27 June, 1990), edited by M. Cveti\v{c} and P. Langacker
(World Scientific, Singapore, 1991)}
\def \waw{XXVIII International Conference on High Energy
Physics, Warsaw, July 25--31, 1996}
\def \yaf#1#2#3#4{Yad. Fiz. {\bf#1}, #2 (#3) [Sov. J. Nucl. Phys. {\bf #1},
#4 (#3)]}
\def \zhetf#1#2#3#4#5#6{Zh. Eksp. Teor. Fiz. {\bf #1}, #2 (#3) [Sov. Phys. -
JETP {\bf #4}, #5 (#6)]}
\def \zpc#1#2#3{Zeit. Phys. C {\bf#1}, #2 (#3)}
\def \zpd#1#2#3{Zeit. Phys. D {\bf#1}, #2 (#3)}


\begin{thebibliography}{99}

\bibitem{CDFsigb} J. Pursley, talk presented on behalf of the CDF Collaboration
at the 11th International Conference on $B$-Physics at Hadron Machines
(Beauty 2006), 25--29 September 2006, University of Oxford, to be published in
Nucl.\ Phys.\ B (Proc.\ Suppl.);
I. Gorelov, presented on behalf of the CDF Collaboration
at the Second Meeting of the APS Topical Group on Hadron Physics (GHP 2006),
22--24 October 2006, Nashville, TN; CDF Collaboration, public web page \\
{\tt http://www-cdf.fnal.gov/physics/new/bottom/060921.blessed-sigmab/}.

\bibitem{Jenkins:1996de} E.~Jenkins,
  Phys.\ Rev.\ D {\bf 54}, 4515 (1996); {\it ibid.} {\bf 55}, 10 (1997).

\bibitem{Karliner:2003sy} M.~Karliner and H.~J.~Lipkin,
  arXiv:hep-ph/0307243 (unpublished).

\bibitem{earlier}
  E.~Bagan, M.~Chabab, H.~G.~Dosch and S.~Narison,
  Phys.\ Lett.\ B {\bf 278}, 367 (1992); B {\bf 287}, 176 (1992);
  S.~Capstick and N.~Isgur,
  Phys.\ Rev.\ D {\bf 34}, 2809 (1986);
  R.~Roncaglia, D.~B.~Lichtenberg and E.~Predazzi,
  Phys.\ Rev.\ D {\bf 52}, 1722 (1995);
  N.~Mathur, R.~Lewis and R.~M.~Woloshyn,
  Phys.\ Rev.\ D {\bf 66}, 014502 (2002).

\bibitem{Ebert:2005xj} D.~Ebert, R.~N.~Faustov and V.~O.~Galkin,
  Phys.\ Rev.\ D {\bf 72}, 034026 (2005).

\bibitem{Acosta:2005mq} D.~Acosta {\it et al.} [CDF Collaboration],
  Phys.\ Rev.\ Lett.\ {\bf 96}, 202001 (2006).

\bibitem{Gasiorowicz:1981jz} S.~Gasiorowicz and J.~L.~Rosner,
  Am.\ J.\ Phys.\ {\bf 49}, 954 (1981).

\bibitem{Kwong:1987ak}W.~Kwong, P.~B.~Mackenzie, R.~Rosenfeld and J.~L.~Rosner,
  Phys.\ Rev.\ D {\bf 37}, 3210 (1988).

\bibitem{Yao:2006px} W.-M. Yao {\it et al.} [Particle Data Group], J. Phys.\ G
{\bf 33}, 1 (2006).

\bibitem{Rosner:1998zc} J.~L.~Rosner,
  Phys.\ Rev.\ D {\bf 57}, 4310 (1998).

\bibitem{DeRujula:1975ge} A.~De R\'ujula, H.~Georgi and S.~L.~Glashow,
  Phys.\ Rev.\ D {\bf 12}, 147 (1975).

\bibitem{Hwang:1986ee} W.~Y.~P.~Hwang and D.~B.~Lichtenberg,
  Phys.\ Rev.\ D {\bf 35}, 3526 (1987).

\bibitem{Capstick:1987cw} S.~Capstick,
  Phys.\ Rev.\ D {\bf 36}, 2800 (1987).

\end{thebibliography}
\end{document}